\documentclass[prl,twocolumn,showpacs,preprintnumbers,amsmath,amssymb]{revtex4}

\usepackage{graphicx}% Include figure files
\usepackage{dcolumn}% Align table columns on decimal point
\usepackage{bm}% bold math

\begin{document}

\preprint{PRL/APS}

\title{Coexisting order and disorder hidden in a quasi-two-dimensional frustrated magnet}

\author{K.~Iida$^{1}$}
\author{S.-H.~Lee$^{1}$}\email{shlee@virginia.edu}
\author{S.-W.~Cheong$^{2}$}

\affiliation{$^1$Department of Physics, University of Virginia, Charlottesville, Virginia 22904, USA}
\affiliation{$^2$Department of Physics and Astronomy, Rutgers University, Piscataway, New Jersey 08854, USA}
\date{\today}

\begin{abstract}
Frustrated magnetic interactions in a quasi-two-dimensional \textless111\textgreater\ slab of pyrochlore lattice were studied. For uniform nearest neighbor (NN) interactions, we show that the complex magnetic problem can be mapped onto a model with two independent degrees of freedom, tri-color and binary sign. This provides a systematic way to construct the complex classical spin ground states with collinear and coplanar bi-pyramid spins. We also identify `partial but extended' zero-energy excitations amongst the ground states. For nonuniform NN interactions, the coplanar ground state can be obtained from the collinear bi-pyramid spin state by collectively rotating two spins of each tetrahedron with an angle, $\alpha$, in an opposite direction. The latter model with $\alpha \sim 30^\circ$ fits the experimental neutron data from SCGO well. 
\end{abstract}

\pacs{Valid PACS appear here}% PACS, the Physics and Astronomy
                             % Classification Scheme.

%\keywords{}

\maketitle
In ordinary magnets, when temperature is lowered, the spins freeze into a long-range ordered state or a spin solid. Some magnets however do not order even at low temperatures~\cite{Villain,Balents,Ramirez1,Bramwell,Ramirez2,Gingras,Gardner,Booth}. The simplest examples are a triangle of three antiferromagnetic (AFM) spins and a tetrahedron of four AFM spins. For both systems, any spin configuration with total zero spin can be a ground state. When such triangles (tetrahedra) are arranged in a two-dimensional (three-dimensional) corner-sharing network or kagome (pyrochlore) lattice, there is an infinite way of covering the entire lattice with the total-zero-spin building blocks. As a result, instead of ordering at low temperatures, the kagome and pyroclore antiferromagnets remain in a spin liquid state~\cite{Moessner1,Moessner2,Canals}. A hallmark of the frustration-driven spin liquid is the existence of local zero energy excitation modes that continuously connect their degenerate ground states in the phase space of spin configuration and energy; the so-called weather-vane mode for the kagome~\cite{Chandra,Delft} and the hexagonal mode for the pyrochlore~\cite{SHL1,Conlon} antiferromagnets.

\begin{figure}[t]
\begin{center}
\includegraphics[width=8.42cm]{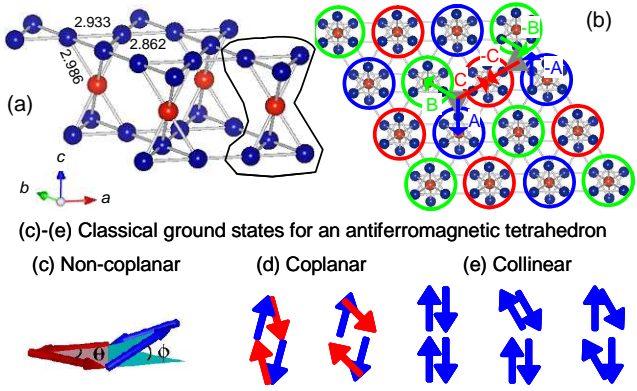}
\caption{\label{Fig:1}
(color online).
(a) The (111)-slab of pyrochlore lattice made up by a kagome(blue spheres)-triangular(red spheres)-kagome tri-layer realized by the magnetic Cr$^{3+}$ ions in SCGO. The numbers are the bond lengths in \AA. The blue and red spheres represent kagome and triangular sites, respectively. (b) The lattice is projected on the $ab$-plane. The red, blue, green arrows represent the three spin directions of a 120$^\circ$ configuration, either (A,B,C) or (-A,-B,-C). (c) Non-coplanar, (d) coplanar, and (e) collinear classical ground states for an AFM tetrahedron. In (e), the first figure represents a collinear state and the other two represent the possible zero-energy modes.
}
\end{center}
\end{figure}

Some frustrated magnets, however, exhibit non-conventional spin-glass behaviors; field-cooled and zero-field-cooled hysteresis in the bulk susceptibility\cite{Ramirez2}, and static short-range spin correlations but with a strongly momentum-dependent structure factor in neutron scattering\cite{SHL3}. Among them, SrCr$_{9p}$Ga$_{12-9p}$O$_{19}$ [SCGO($p$)]~\cite{Obradors,Broholm,SHL2,SHL3,Ramirez3,Keren,Mendels,Limot,Mutka} and Ba$_2$Sn$_2$ZnGa$_3$Cr$_7$O$_{22}$ (BSZGCO)~\cite{Mutka,Hagemann,Bono1,Bono2} are particularly interesting because in both systems the magnetic Cr$^{3+}$ ($3d^3$; $s=3/2$) ions form a kagome-triangular-kagome tri-layer.\cite{dilution} 
Due to the ligand environment and the electronic orbitals of the Cr$^{3+}$ ions \cite{SHL2}, the spin Hamiltonian can be described by $\mathcal{H}=J\sum_\text{k}\mathbf{S}_i\cdot\mathbf{S}_j+J'\sum_\text{k-t}\mathbf{S}_i\cdot\mathbf{S}_j$ where the first sum is over the nearest neighbor (NN) bonds between the kagome spins and the second sum is over the bonds between the kagome and triangular spins [Fig.~\ref{Fig:1}(a)]. The different values of $J$ and $J'$ are due to their different bond lengths. Since the discovery of SCGO more than two decades ago~\cite{Obradors}, understanding the origin of the non-conventional spin glass behavior has been a challenging issue. 

In this paper, by mapping the magnetic interaction problem onto a model with two independent degrees of freedom, tri-color and binary sign, we present a methodical way to construct the complex classical spin ground states with collinear and coplanar bi-pyramidal spins for uniform and nonuniform NN interactions. We also identify `partial but extended' zero-energy excitations amongst the ground states that are qualitatively different from the `local' zero-energy excitations found in spin liquid states of other frustrating magnets. By comparing the resulting theoretical magnetic scattering to the experimental neutron intensities obtained from single crystals of SCGO($p=0.67)$, the ratio of $J'/J\sim0.70(15)$ was obtained, which is reasonable for SCGO. We argue that a topological spin glass might be the ground state of the magnetic lattice.

Elastic neutron scattering measurements on single crystals of SCGO($p=0.67$) with total mass of $\sim300$~mg were performed on the cold neutron triple-axis spectrometer SPINS at the NIST Center for Neutron Research. Energy of neutrons was fixed to $E_\text{i}=5$~meV with an energy resolution of 0.25~meV. A horizontally focusing analyzer with seven PG(002) blades was utilized to enhance the intensity of scattered neutrons. The background and nuclear contributions to the detector count rate have been measured at $T=20$~K that is above its spin glass transition, $T_\text{g}=4.5$~K, and subtracted from the 1.6~K data. 

Let us first consider the perfect quasi-two-dimensional lattice with $J'=J$. 
We consider two corner-sharing tetrahedra or a bi-pyramid with a triangular base circled by a solid line in Fig.~\ref{Fig:1}(a) as a `molecular' unit. As shown in Fig.~\ref{Fig:1}(b), the bi-pyramids form a triangular superlattice. This is crucial because triangles and tetrahedra behave quite differently; the triangle favors a 120$^\circ$ spin configuration [Fig.~\ref{Fig:1}(b)] while the tetrahedron favors any configuration with total zero spin [Figs.~\ref{Fig:1}(c)--\ref{Fig:1}(e)]. The challenge is then to find spin configurations where every tetrahedron within the bi-pyramids and the triangles that link neighboring bi-pyramids [highlighted by grey triangles in Fig.~\ref{Fig:1}(b)] satisfy their different AFM constraints. One can easily see that there will be an infinite number of such configurations. One class of spin configurations that we will focus on first is those where spins within the bi-pyramids are collinear. This approach can be justified by the following argument. 
Let us consider the zero-energy excitations possible for non-coplanar, coplanar, and collinear spin configurations for an AFM tetrahedron. For the non-coplanar state, if one spin rotates then the other three spins must rotate to keep the total spin to be zero [Fig.~\ref{Fig:1}(c)]. For the coplanar state with two pairs of antiparallel spins, on the other hand, if one spin rotates, only its counterpart forming one pair can rotate by the same angle to keep the total spin zero without moving the other pair [Fig.~\ref{Fig:1}(d)]. Such zero energy excitations involving only a fraction (half) of spins would favor a coplanar ground state over a non-coplanar state. For the collinear case, the two pairs are along the same direction, and thus there are two choices forming a pair [Fig.~\ref{Fig:1}(e)]. Thus for a single bi-pyramid the collinear structure is favored by entropy. Of course, the coplanar state has also extensive entropy, and cannot be ignored. After finding the possible collinear bi-pyramid spin states for the hybrid lattice, we will show that the coplanar bi-pyramid spin states can be generated from the collinear states. 

To find possible collinear bi-pyramid spin states, let us start by assigning, for a linking triangle a 120$^\circ$ spin configuration with three spins (A, B, C) colored in red, blue, and green, respectively [Fig.~\ref{Fig:1}(b)]. Since the three spins belong to three different bi-pyramids, the three different colors can be assigned to the bi-pyramids to show that all seven spins within each bi-pyramid point either parallel or antiparallel to the assigned spin direction. Since the bi-pyramids are connected in a triangular superlattice, the inter-bi-pyramid interactions for the collinear bi-pyramids can only be minimized when the colors form a long-range $\sqrt{3}\times\sqrt{3}$ color structure [see the colored circles in Fig.~\ref{Fig:1}(b)]. The remaining task is to find the internal spin state of the collinear bi-pyramid. This can be done simply by assigning plus or minus signs to the seven spins forming a bi-pyramid with the constraint that each tetrahedron must have two plus and two minus signs. When such bi-pyramids are arranged in the triangular superlattice, only one constraint has to be satisfied: the linking triangle must have a 120$^\circ$ spin configuration, i.e., either (A,B,C) or (-A,-B,-C), leading to a strict constraint of ferro-sign bonds for the triangles [Fig.~\ref{Fig:1}(b)]. The spin degrees of freedom are thus mapped onto two different degrees of freedom, the long-range ordered tri-color and the disordered binary sign. 

\begin{figure}[t]
\begin{center}
\includegraphics[width=8.54cm]{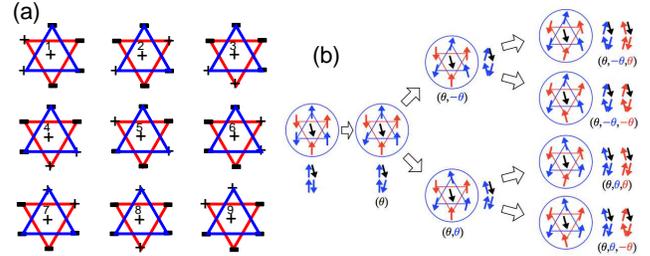}
\caption{\label{Fig:2} 
(color online). 
(a) Nine possible sign states for a bi-pyramid where each tetrahedron has total zero spin. (b) Collective zero-energy excitations for a bi-pyramid involving all seven spins and satisfying the total zero spin constraint. 
}
\end{center}
\end{figure}

\begin{figure}[b]
\begin{center}
\includegraphics[width=8.54cm]{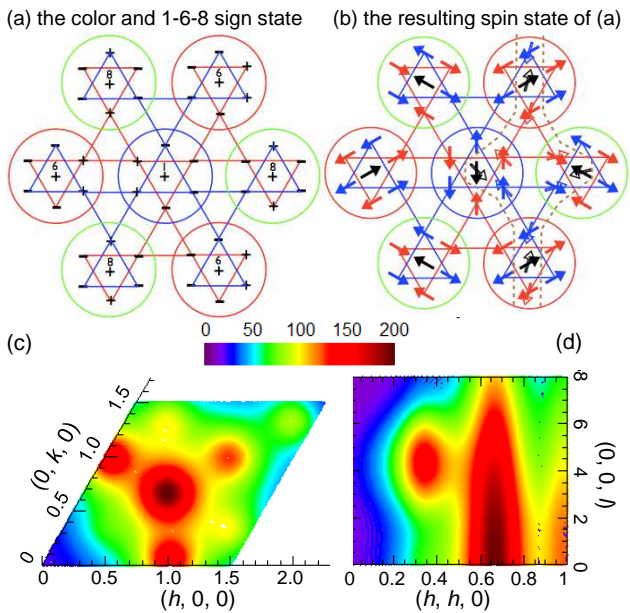}
\caption{\label{Fig:3}
(color online).
(a) A sign state for the triangular lattice of bi-pyramids that has a long range $\sqrt{3}\times\sqrt{3}$ structure. (b) Filled arrows represent spins in a collinear bi-pyramid spin state constructed by the color and the 1-6-8 sign state shown in (a). Open arrows are explained in the text. (c), (d) The calculated neutron scattering intensities for $J'=J$ in the $(hk0)$ and $(hhl)$ planes.
}
\end{center}
\end{figure}

Each collinear bi-pyramid has 18 possible sign states, nine of which are shown in Fig.~\ref{Fig:2}(a) and the other nine can be generated by a sign flip. 
%These 9 sign states are furthermore inter-related by a three-fold rotation. 
%Thus, tiling the bi-pyramids over the triangular superlattice can be considered as an 18-%state Potts model. 
Thus, for non-interacting $N$ number of bi-pyramids there are $18^N$ possible sign states. The ferro-sign bond constraint, however, severely limits the options. To see this, let us assign a sign state (say state 1) to a central bi-pyramid, and determine all possible sign states for the 6 nearest neighboring bi-pyramids of a hexagonal ring. It is tedious but straightforward to see that there are only 111 number of sign states allowed by the ferro-sign bonds. When the 12 bi-pyramids on the next large hexagonal ring are included, there are 13238 possible sign states for the 18 neighboring bi-pyramids. There is however only one sign state that has a long range order, and is formed by the three sign states, 1, 6 and 8, as shown in Fig.~\ref{Fig:3}(a). It is easy to find that there are three similar long-range ordered sign states formed by the sign states 1 to 9; 1-6-8 [Fig.~\ref{Fig:3}(a)], 2-4-9, and 3-5-7. There is another set of such three states that can be obtained by the sign flip symmetry. Once the sign state is determined for the lattice, the actual spin state can be easily constructed by imposing the color state on the sign state, as shown in Figs.~\ref{Fig:3}(a) and \ref{Fig:3}(b).

Let us now turn to the coplanar bi-pyramid spin states.
% that cannot be ignored because of their extensive entropy. 
A coplanar state can be generated %for a bi-pyramid 
from a collinear state by rotating the 7 spins in several collective ways (see Fig.~\ref{Fig:2}(b)). For the entire lattice, because of the color and ferro-sign bond constraints, three angles with the same magnitude, $(\theta,\pm\theta,\pm\theta)$, are sufficient to generate a long range ordered coplanar state.
% over the entire hybrid lattice. 
An arbitrary value of $\theta$ corresponds to a coplanar state which leads to macroscopically degenerate coplanar states. The collinear spin state and its resulting coplanar states are continuously connected with each other in the spin-energy phase space, and the excitations among them realize the zero-energy excitations involving all spins - `global' spin zero-energy excitations. Different types of zero-energy excitations are also possible for the long range ordered collinear and coplanar bi-pyramid spin states. As illustrated in Figs.~\ref{Fig:1}(d) and \ref{Fig:1}(e), only one pair of the antiparallel spins of a coplanar or collinear tetrahedron can rotate without moving the other pair. This kind of excitations leads to `partial but extended' spin zero-energy excitations. An example is the excitations of an uneven spaghetti shape that propagates along only one-direction in the $ab$-plane, as illustrated by the open arrows and the dashed line in Fig.~\ref{Fig:3}(b).

If a real material realizes the model of $\mathcal{H}$ with $J'=J$, upon cooling the spins would freeze into numerous finite size domains with every collinear and coplanar bi-pyramid spin state. We generated all possible long-range coplanar states by the angle, $\theta$, with a step of $5^\circ$ for the `global' spin zero-energy motion starting from the aforementioned three long range collinear states, and calculated the square of the magnetic structure factor, $|F_\text{M}(\mathbf{Q})|^2$, for each state. For long range order, $I(\mathbf{Q})\propto\sum_{\mathbf{Q}_\text{M}}|F_\text{M}(\mathbf{Q})|^2\delta^2(\mathbf{Q}-\mathbf{Q}_\text{M})$, with two dimensional reciprocal lattice vectors $\mathbf{Q}_\text{M}=(n/3,m/3,0)$ where $n$ and $m$ are integers. For short range order, however, the two-dimensional delta function $\delta^2(\mathbf{Q}-\mathbf{Q}_\text{M})$ is replaced by the two-dimensional lorentzian $1/[\kappa^2+(\mathbf{Q}_{ab}-\mathbf{Q}_\text{M})^2]$ where $\mathbf{Q}_{ab}$ is the $ab$-component of $\mathbf{Q}$ and $\kappa$ is the inverse of the in-plane correlation length, $\xi$, $\kappa=2\pi/\xi$. 
Fig.~\ref{Fig:3}(c) and (d) show the calculated $I_\text{ave}(\mathbf{Q})$ with $\xi\sim20$~$\text{\AA}$. %that is $\sim7$ times longer than the NN distance between spins. 
The $I_\text{ave}(\mathbf{Q})$ produces strong broad scattering at around $(2/3,2/3,l)$ and $(1/3,1/3,l)$ in the $(hhl)$ plane, and $(2/3,2/3,0)$ in the $(hk0)$ plane. It also produces additional peaks at $(1,0,0)$ and $(1,1,0)$ in the $(hk0)$ plane, and along $(1,1,l)$ in the $(hhl)$ plane.

\begin{figure}[t]
\begin{center}
\includegraphics[width=7.77cm]{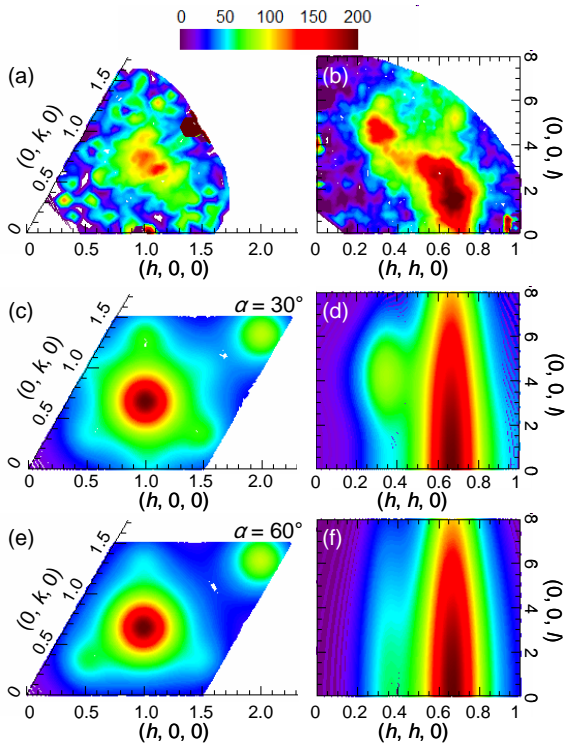}
\caption{\label{Fig:5} 
(color online).
(a) and (b) The experimental magnetic neutron scattering intensities in the $(hk0)$ and $(hhl)$ planes. (c)-(f) The calculated elastic magnetic scattering intensities with $\alpha=30^\circ$ [(c),(d)] and with $\alpha=60^\circ$ [(e),(f)].
}
\end{center}
\end{figure}

Fig.~\ref{Fig:5}(a) and (b) show contour maps of magnetic neutron scattering intensities obtained from single crystals of SCGO($p=0.67$). The experimental data has similarities and differences with the calculated intensities for $J'=J$; their similarity is the strong broad peaks at $(2/3,2/3,0)$ in the $(hk0)$ plane and along the (1/3,1/3,$l$) an (2/3,2/3,$l$) directions, while their difference is the lack of scattering at the integer $Q$-points in the $(hk0)$ plane and along the $(1,1,l$) direction in the $(hhl)$ plane. The discrepancy comes from the fact that SCGO does not realize the perfect hybrid lattice with $J'=J$. Instead the lattice is distorted to yield different coupling constants for the in-plane and out-of-plane interactions as shown in Fig.~\ref{Fig:1}(a).

\begin{figure}[t]
\begin{center}
\includegraphics[width=8.42cm]{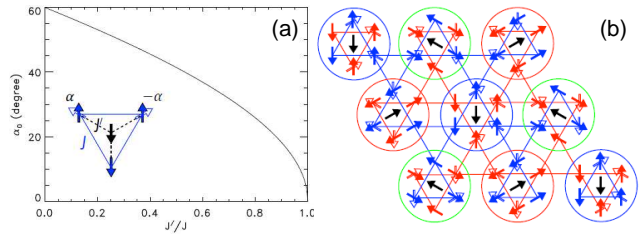}
\caption{\label{Fig:6}
(color online).
(a) The optimal $\alpha_0$ is plotted as a function of $J'/J$ for the tetrahedron. (b) The long-range ordered conventional $\sqrt{3}\times\sqrt{3}$ structure constructed for $J'=0$, explained in the text. A similar rotation with $(-60^\circ,60^\circ)$ will lead to the $q=0$ structure.
}
\end{center}
\end{figure}

 While the tricolor-sign state, i.e. the collinear bipyramid spin state, can no longer be the ground state for the non-uniform exchange interactions, here we show that the collinear state can be a good reference point to construct the ground state for the non-uniform case. Consider a collinear state for a single tetrahedron shown as the filled arrows in the inset of Fig.~\ref{Fig:6}(a). The ground state for $J'<J$ can be obtained by rotating two kagome spins (unfilled arrows) that are antiparallel to the triangular spin in opposite direction, $(\alpha,-\alpha)$. The magnetic energy of the resulting spin configuration becomes $E_\alpha=J[(-2\text{cos}\alpha+\text{cos}2\alpha)+J'/J(1-2\text{cos}\alpha)]$ that is lower than the energy of the collinear state $E_{\alpha=0}=-J-J'$. The optimal angle, $\alpha_0$, for a given value of $J'/J$ can be obtained by differentiating $E_\alpha$ with respect to $\alpha$, which yields $J'/J=2\text{cos}\alpha_0-1$. Fig.~\ref{Fig:6}(a) shows the $\alpha_0$ as a function of $0\le J'/J\le1$. For $J'=J$, $\alpha_0=0$ represents the collinear state, while for $J'=0$, $\alpha_0=60^\circ$ represents the conventional 120$^\circ$ noncollinear spin configuration. There are only two ways to rotate the two kagome spins; $(\alpha,-\alpha)$ or $(-\alpha,\alpha)$. When such bipyramids are arranged to form the triangular lattice of bipyramids as realized in SCGO, due to the strict antiferromagnetic constraint for the linking triangles, each rotation leads to a single long-range state. A resulting long-range ordered non-collinear coplanar bipyramid states with $(\alpha,-\alpha)$ is shown in Fig.~\ref{Fig:6}(b) for the special angle of $\alpha_0=60^\circ$ for $J'=0$, which is the conventional $\sqrt{3}\times\sqrt{3}$ structure for kagome planes.

We considered the three long range ordered collinear bi-pyramid states described above, and performed the collective $\alpha-$rotation towards the conventional $\sqrt{3}\times\sqrt{3}$ structure. In each case, the square of the magnetic structure factor, $|F(\mathbf{Q})|^2$, was averaged over three crystallographic domains, and in the end, all three cases were added. The same two-dimensional lorentzians used in the uniform $J$ case to account for the short range order were multiplied to $|F(\mathbf{Q})|^2$ to obtain the magnetic neutron scattering intensities, $I(\mathbf{Q})$, for various values of $\alpha$, as shown in Figs.~\ref{Fig:5}(c)--\ref{Fig:5}(f) for $\alpha=30^\circ$ and $60^\circ$. When $\alpha$ increases, the intensities at the $q=0$ points such as $(1,0,0)$, $(0,1,0)$, and $(1,1,0)$ in the $(hk0)$ plane decrease, as well as those at $(1/3,1/3,l\sim4.2)$ and along the $(1,1,l)$-direction in the $(hhl)$ plane. For $\alpha=60^\circ$, the $(1/3,1/3,l\sim4.2)$ peak almost disappears, which is inconsistent with the data. This means that $\alpha<60^\circ$ for SCGO. Because of the $\sim$ 30\% nonmagnetic defects in the single crystals of SCGO($p=0.67$), it is not possible to quantitatively compare the data to the theoretical model. However, qualitative comparison with the calculated $I(\mathbf{Q})$ with $\alpha\sim30(10)^\circ$ reproduces the salient features of the data, as shown in Fig.~\ref{Fig:5}. The $\alpha\sim30(10)^\circ$ corresponds to $J'/J\sim0.70(15)$, which is a reasonable value for SCGO considering their bond lengths.

In summary, we have presented a systematic approach involving a mapping into two degrees of freedom, to dealing with the complex magnetic interactions in the quasi-two-dimensional \textless111\textgreater\ slab of pyrochlore lattice. 
%By comparing the resulting theoretical magnetic scattering to the experimental neutron intensities obtained from single %crystals of SCGO($p=0.67)$, the ratio of $J'/J\sim0.70(15)$ was obtained, which is reasonable for SCGO. 
%Starting from the simple mapping in the uniform $J$ case, 
This method can be used to obtain the ground states not only for the uniform NN $J$ case but also for the nonuniform $J$ case. Both uniform and nonuniform $J$ cases have degenerate ground states, that provide a topological argument for the non-conconventional spin glass state of this system. We expect such spin glassy behaviors to be enhanced in the uniform $J$ case because of the existence of the unique `partial' spin zero energy excitation modes.

\begin{acknowledgments}
We thank C. Broholm and A. B. Harris for discussions. Work at UVa was supported by the U.S. Department of Energy, Office of Basic Energy Sciences, Division of Materials Sciences and Engineering under Award DE-FG02-10ER46384. This work utilized facilities supported in part by the National Science Foundation under Agreement No. DMR-0944772.
\end{acknowledgments}


\begin{thebibliography}{40}

\bibitem{Villain}%1
J. Villain \textit{et} \textit{al.}, 
Z. Phys. B \textbf{33}, 31 (1979).

\bibitem{Balents}%2
L. Balents \textit{et} \textit{al.}, 
Nature (London) \textbf{464}, 199 (2010).

\bibitem{Ramirez1}%3
A. P. Ramirez \textit{et} \textit{al.}, 
Nature (London) \textbf{399}, 333 (1999).

\bibitem{Bramwell}%4
S. T. Bramwell \textit{et} \textit{al.}, 
Science \textbf{294}, 1495 (2001).

\bibitem{Ramirez2}%5
A. P. Ramirez \textit{et} \textit{al.}, 
Phys. Rev. Lett. \textbf{64}, 2070 (1990).

\bibitem{Gingras}%6
M. J. P. Gingras \textit{et} \textit{al.}, 
Phys. Rev. Lett. \textbf{78}, 947 (1997).

\bibitem{Gardner}%7
J. S. Gardner \textit{et} \textit{al.}, 
Phys. Rev. Lett. \textbf{83}, 211 (1999).

\bibitem{Booth}%8
C. H. Booth \textit{et} \textit{al.}, 
Phys. Rev. B \textbf{62}, R755 (2000).

\bibitem{Moessner1}%9
R. Moessner \textit{et} \textit{al.}, 
Phys. Rev. Lett. \textbf{80}, 2929 (1998).

\bibitem{Moessner2}%10
R. Moessner \textit{et} \textit{al.}, 
Phys. Rev. B \textbf{58}, 12049 (1998).

\bibitem{Canals}%11
B. Canals \textit{et} \textit{al.}, 
Phys. Rev. Lett. \textbf{80}, 2933 (1998).

\bibitem{Chandra}%12
P. Chandra \textit{et} \textit{al.}, 
J. Phys. I France \textbf{3}, 591 (1993).

\bibitem{Delft}%13
J. von Delft \textit{et} \textit{al.}, 
Phys. Rev. B \textbf{48}, 965 (1993).

\bibitem{SHL1}%14
S.-H. Lee \textit{et} \textit{al.}, 
Nature (London) \textbf{418}, 856 (2002).

\bibitem{Conlon}%15
P. H. Conlon \textit{et} \textit{al.}, 
Phys. Rev. B \textbf{81}, 224413 (2010).

\bibitem{SHL3}%19
S.-H. Lee \textit{et} \textit{al.}, 
Europhys. Lett. \textbf{35}, 127 (1996).

\bibitem{Obradors}%16
X. Obradors \textit{et} \textit{al.}, 
Solid State Comm. \textbf{65}, 189 (1988).

\bibitem{Broholm}%17
C. Broholm \textit{et} \textit{al.}, 
Phys. Rev. Lett. \textbf{65}, 3173 (1990).

\bibitem{SHL2}%18
S.-H. Lee \textit{et} \textit{al.}, 
Phys. Rev. Lett. \textbf{76}, 4424 (1996).

\bibitem{Ramirez3}%20
A. P. Ramirez \textit{et} \textit{al.}, 
Phys. Rev. Lett. \textbf{84}, 2957 (2000).

\bibitem{Keren}%21
A. Keren \textit{et} \textit{al.}, 
Phys. Rev. Lett. \textbf{84}, 3450 (2000).

\bibitem{Mendels}%22
P. Mendels \textit{et} \textit{al.}, 
Phys. Rev. Lett. \textbf{85}, 3496 (2000).

\bibitem{Limot}%23
L. Limot \textit{et} \textit{al.}, 
Phys. Rev. B \textbf{65}, 144447 (2002).

\bibitem{Mutka}%24
H. Mutka \textit{et} \textit{al.}, 
Phys. Rev. Lett. \textbf{97}, 047203 (2006).

\bibitem{Hagemann}%25
I. S. Hagemann \textit{et} \textit{al.}, 
Phys. Rev. Lett. \textbf{86}, 894 (2001).

\bibitem{Bono1}%26
D. Bono \textit{et} \textit{al.}, 
Phys. Rev. Lett. \textbf{92}, 217202 (2004).

\bibitem{Bono2}%27
D. Bono \textit{et} \textit{al.}, 
Phys. Rev. Lett. \textbf{93}, 187201 (2004).

\bibitem{dilution}
These systems have vacancies in the magnetic lattice. Several  studies using different experimental techniques, such as NMR\cite{Limot}, specific heat\cite{Ramirez3} and neutron scattering\cite{Broholm, SHL3}, however have shown that the observed spin freezing is not due to the defects but is intrinsic to the trilayer antiferromagnet. 

\end{thebibliography}
\end{document}